\documentstyle[prl,aps,multicol,psfig]{revtex}
\draft
\begin{document}
\title{ Chaos and Exponentially 
 Localized Eigenstates
 in Smooth Hamiltonian Systems}
\author{    
M. S. Santhanam \cite{byline}, V. B. Sheorey \cite{byline}, and A. Lakshminarayan \cite{byline} }
\address{ Physical Research Laboratory,
Navrangpura, Ahmedabad - 380 009, India}
\date{\today}

\maketitle

\begin{abstract}
We present numerical evidence to show that the  wavefunctions of 
smooth classically chaotic Hamiltonian systems
scarred by certain simple
periodic orbits are exponentially localized in the space of
unperturbed basis states. The degree of localization, as measured by the 
information entropy, is shown to be correlated with the  local phase
space structure around the scarring orbit; indicating sharp localization 
when the orbit undergoes a pitchfork bifurcation and loses stability. 
\end{abstract}


\begin{multicols}{2}
It is now well-known that  classical periodic
orbits have an enduring influence on the quantum mechanics and the semiclassics
of classically chaotic quantum systems. The knowledge of all the 
isolated periodic orbits of such a system allows us to semiclassically
estimate the
eigenvalues through the application of Gutzwiller's trace formula \cite{gut}.
Heller \cite{hel} has given theoretical arguments to show that for
the classically chaotic quantum systems the eigenfunctions will show 
density enhancements, called scars, along the least unstable periodic 
orbits. Such localized probability
density structures of the wavefunctions and their correspondence to the
underlying periodic orbits are widely reported for many classically chaotic
quantum systems, such as the hydrogen atom in a magnetic field \cite{muw}.
The effect of localized states, in a quantum
system with predominant classical chaos, has been experimentally
observed using tunnel-current spectroscopy in semiconductor heterostructures
\cite{wil}.
The role played by the bifurcation properties of periodic orbits are also of
considerable importance. The effect of orbit bifurcation  
is observed experimentally in the spectra of atoms in external
fields \cite{mde,bif}.

The first  numerical evidence for wavefunction localization in systems
with classical chaos came from the studies of Bunimovich billiards
\cite{hel,mck}. For the
kicked rotor, Grempel et. al. \cite{gpf} have shown that the localization 
in momentum space is
exponential, similar to Anderson localization in the case
of charged particle dynamics in a series of potential wells with random
depths.  In smooth potentials like the coupled oscillators, 
adiabatic methods have been widely applied to predict some eigenvalues but
the construction of adiabatic wavefunctions still remains an open problem.
Polavieja et. al. \cite{pbb} have used wavepacket propagation techniques 
to construct
wavefunctions highly localized on a given classical periodic orbit.
A qualitative study of
the effect of pitchfork bifurcation of simple orbits on the eigenfunctions
has also been reported \cite{aez}.
In this Letter, we explore further the connection between certain simple
classical periodic orbits, their stability and the localization of
the quantum wavefunctions scarred by them, using coupled nonlinear
oscillators.  In such systems, even in the regions of large scale 
classical chaos, certain simple
periodic orbits with  short time periods and high stability are
known to scar a series of WKB-like states in the spectrum \cite{ssa} and
recently it has been observed that these states affect the
eigenvalue spacing distributions \cite{del}.
In particular, our numerical evidence for smooth potentials indicates
that such states, scarred by simple periodic orbits, are exponentially
localized in the space of unperturbed
basis states. 

In the general analytical framework of scarring as developed by
Berry \cite{ber}, the spectral Wigner function averaged over a small 
energy scale is shown to be influenced by isolated periodic orbits
semiclassically. The stability of the periodic orbits is shown to 
affect the scar weight and scar amplitude significantly. The analysis 
remains limited to averaged Wigner functions and is not valid
at the points of bifurcations where  Gutzwiller's density of 
states formula breaks down, leading to predictions of either infinite 
scar weights or scar amplitudes. In this context we will show that gross
measures of individual wavefunction localization, such as entropy, 
are strongly correlated 
with the periodic orbit's stability oscillations (with a parameter). 
When the orbit loses stability we will find that the entropy  is also
low, with the point of bifurcation being approximately the point of 
a local minimum in the entropy (again as a function of a parameter). 
We will note that this does not always coincide with the points at which the 
Berry formula predicts an infinite scar amplitude.

We study smooth Hamiltonian systems of the form,
\begin{equation}
\label{Ham}
H(x,y,p_{x},p_{y};\alpha) = \frac{p_{x}^2}{2\, m} + \frac{p_{y}^2}{2 \, m} 
+ V(x,y;\alpha)
\end{equation}
whose classical dynamics can be regular or chaotic depending on the
value of the parameter $\alpha$. The coupled quartic oscillator given
by the homogeneous potential $ V(x,y;\alpha) = x^4 + y^4 + \alpha x^2 y^2 $, 
is used below 
with $m=1/2$. This
system is integrable for $\alpha = 0, 2$ and $6$ and
exhibits increasing chaos as $\alpha$ is increased beyond $ 6 $. 
The presence of a ``channel'' in this potential leads to trapping of the
particle in a motion along the one dimensional channel periodic orbit 
$(x,y=0,p_{x},p_{y}=0)$,
which has a short time-period and interesting stability properties.
The stability of the channel periodic orbit as measured by 
$\mbox{Tr} J(\alpha)$, where $J(\alpha)$ is the monodromy matrix, 
displays bounded oscillations as a function of $\alpha$ and for this
simple periodic orbit of the Hamiltonian in Eq.\ (\ref{Ham}) the analytical expression for 
$\mbox{Tr} J(\alpha)$, due to Yoshida \cite{yos}, is given by,
\begin{equation}
\label{Yoshi}
\mbox{Tr}\;J(\alpha)= 2 \sqrt{2} \cos( \frac{\pi}{4} \sqrt{1+4 \alpha}),
\end{equation}

The monodromy matrix refers to the ``half Poincar\'{e} map'' \cite{mde},
which is the usual Poincar\'{e} map without the restriction of positive 
momentum on the intersection. We note that this is the appropriate monodromy
matrix in the semiclassical analysis as well due to the fact that we are 
restricting the quantum spectrum to the $A_{1}$ irreducible representation
of the $C_{4v}$ point group which the Hamiltonian  possesses.
The important inference from Eq.\ (\ref{Yoshi}) is that this channel 
orbit changes 
stability when  $\alpha=n(n+1)$, where $n$ is an integer,
through a pitchfork bifurcation. Another interesting property we have
recently found is that the Poincar\'{e}
section around channel periodic orbit locally scales with respect
to the parameter $\alpha$ for all one-parameter homogeneous two-dimensional
Hamiltonian systems, which 
includes the coupled quartic oscillator and the 
scaling exponents depend simply on the degree of homogeneity of the 
potential \cite{ass}. In this letter, we will focus our attention on this 
channel periodic orbit and its influence on the quantum wavefunctions.

The quantization of the Hamiltonian in Eq.\ (\ref{Ham}) 
is by numerically solving the Schr\"{o}edinger equation
in the basis of the eigenfunctions of the corresponding unperturbed problem,
namely the $\alpha=0$ case in Hamiltonian Eq.\ (\ref{Ham}).
The actual basis states $\psi(x,y)$ employed 
are the symmetrised linear combination of the wavefunctions for 
the unperturbed problem \cite{details}.
The
wavefunction for the $m$th state is given by 
$\Psi_{m}(x,y;\alpha) = \sum_{i=1}^{N} a_{m,j} \psi(x,y)$, where
$j$ represents the pair of even integers $(n_{1},n_{2})$
corresponding to the quantum numbers of two one-dimensional states
that makes up the basis state and $a_{m,j}$ are the expansion
coefficients. The information entropy measure, for the $m$th state
with $M$ components, where $M$ is the dimension of the Hamiltonian
matrix, defined as,
\begin{equation}
{\cal{S}}^{\alpha}_{m} =- \sum_{j=1}^{M} \mid a^{\alpha}_{m,j} \mid ^{2}
                \log \mid a^{\alpha}_{m,j} \mid ^{2}
\end{equation}
is applied to the wavefunction spectrum and we have shown that the
localized states scarred by the channel periodic orbit are identified
by a pronounced dip in the information entropy curve and
they were also visually identified \cite{ssa}.

The Berry-Voros conjecture that the Wigner function of a
typical eigenstate of a quantized chaotic
system is  a microcanonical distribution on the energy shell
is of course flagrantly violated by many scarred states and especially so the
channel localized ones. In \cite{ber} the conjecture  is shown to be 
true semiclassically when there
is a large energy averaging done, while individual isolated orbits enter
as density enhancements as fewer eigenstates are averaged over in the 
spectral Wigner function. The channel orbits are in fact so localized that,
as we discuss below, they are in a sense even exponentially localized.  

In Fig. 1(a) we show the $a_{m,j}$ plotted as a function of $j$ for
one such highly excited state at $\alpha = 90.0$. In Fig. 1(b) $a_{m,j}$
corresponding to a channel localized state at the same value of $\alpha$
is plotted. We immediately recognize that for the channel localized state
very few basis states contribute to building up of the wavefunction, and
we identify the principal contribution to be coming from (N,0),
(N,2), (N,4)... type of basis states, where the even integer $N$ 
refers to the number of quanta of 
excitation for the motion along the channel. 
We observed that
the pattern in Fig. 1(b) is qualitatively  generic for all the channel
localized states. For instance, in Fig. 2 we have the wavefunctions of
localized states for the parameters $\alpha = 88.0$ and $96.0$.

Significantly, the quanta 
$N$ also enters an  adiabatic formula, with constants $b_{0}\approx b_{1}/3$ 
and $b_{1}$,
\begin{equation}
\label{Adiab}
E_{N}(\alpha)=b_{0} \sqrt{\alpha} \left(N + \frac{1}{2}\right)^{1/3}+
b_{1} \left( N + \frac{1}{2} \right)^{4/3}
\end{equation}
to estimate the eigenvalues $E_{N}$ of
the channel localized states. The existence of such a formula for a subclass
of states in a chaotic spectrum is indicative of the special nature of these
states as no systematics has yet been uncovered for scarring.  Although this 
formula does not include the stability of the orbit as a parameter we have 
found that this is most accurate when the orbit is about to lose 
stability and the 
wavefunction is highly localized as noted below. We derived 
 Eq.\ (\ref{Adiab}) as a  
consequence of adiabaticity as developed in \cite{zak}.

Our numerical results show that the channel localized 
states in the unperturbed basis are dominated by exponentially falling peaks
in the quantum number of the motion perpendicular to the direction of 
the channel;  and further that the degree of
localization is related to the stability of the channel periodic
orbit. In Fig. 3 we plot
$\log(a_{m,j(N,n_{2})})$ as a function of $n_{2}$ for the principal peaks corresponding
to the states shown in Fig. 1(b) and Fig. 2 and the fairly good straight 
lines obtained show that the fall is indeed exponential.
The next dominant peak with contributions coming mainly from $(N+2,0), 
(N+2,2), (N+2,4) ....$ basis states also provides an evidence (not shown here)
of exponential localization.
However, for the third peak, corresponding to $(N+4,0)$, the values of
$a_{m,j}$ fall within our accuracy of our calculation and hence although
unequivocal conclusions cannot be drawn we expect the fall to be
exponential.

It has been established that the eigenstates of time-dependent systems,
like the kicked rotor, are exponentially localized but the above result is
the first observation of exponential localization in smooth chaotic systems.
Why not exponential localization for other scarred
states? We believe that the answer lies in the fact that (a) the basis
in which there is exponential localization belongs to the Hamiltonian,
namely Eq.\ (\ref{Ham}) with $\alpha=0$,
for which the scarring orbit is also a valid orbit, (b) The stability of
such an orbit is high, the channel orbit never becomes very unstable,
however large the nonlinearity. For instance the $45^{o}$ straight
line periodic orbits could be exponentially localized in the $45^{o}$ 
rotated unperturbed basis, but this orbit becomes extremely unstable
thereby creating complex states in which other orbits contribute to
the scarred state.

The channel periodic orbit loses stability through a pitchfork
bifurcation ($\mbox{Tr}\;J(\alpha)=-2$) at $\alpha = 90.0$ while giving 
birth to two other stable
orbits. We notice from Fig. 3 that we get the best exponential behavior
at $\alpha = 90.0$ and it progressively moves away from exponential
behavior as we explore the parameter regions in which the channel orbit
also becomes progressively unstable. In Fig. 4 we have  
striking visual evidence for two wavefunctions at two different
$\alpha$ values at which the channel orbit is stable and unstable.
The  wavefunction at $\alpha = 90.0$ is compact and has almost collapsed on 
to the periodic orbit in comparison with the wavefunction at $\alpha = 96.0$.

We calculated the average
information entropy for a particular $\alpha$ by taking the mean
of information entropies,
$
<S_{\alpha}> = \sum_{\sigma} {\cal S}^{\alpha}_{\sigma}/k
$
for a group of $k$ localized states represented by $\sigma$ within some
energy range. The plot of $<S_{\alpha}>$ in Fig. 5 shows that even this 
averaged measure reflects
the trend observed in the stability oscillations of the channel orbit in the 
vicinity of $\alpha = 90.0 $; although the exact minimum of the entropy 
seems to be slightly removed from this point of bifurcation. 
Further evidence of entropy oscillations
is seen in the perturbed oscillator system we present below.

In order to check the validity of our finding on other similar systems, 
we studied the perturbed oscillator, given by the potential $V(x,y)=
x^2/2 + y^2/2 + \beta x^2 y^2 $
where $\beta$ is the
parameter; we have taken $m=1$ and $\beta=0.1$ below. This is a 
non-homogeneous system and the scaled 
parameter is $\epsilon = \beta E$, where $E$ is the energy of the system.
The advantage of non-homogeneity in this system is that now the 
stability oscillations of the channel orbit is with respect to $\epsilon$,
and hence with a fixed value of $\beta$ we can study how the wavefunction
localization is affected by these oscillations. 

The results presented below pertain to the perturbed oscillator, unless
otherwise specified.  
The structure of channel localized wavefunctions
in the unperturbed space does have the generic pattern similar to the 
Fig. 1(b) for the case of coupled quartic oscillator. The principal peak
for the three states, with different scaled parameters 
$\epsilon = 8.03, 19.73$ and $10.65$
corresponding to stable and unstable motion of the channel orbit,
falls exponentially as is evident from Fig. 6, showing the plot of 
$\log(a_{m,j})$ against $j$. Since the scaled parameter $\epsilon$
is a function of the energy $E$, a global picture of the stability of
the channel orbit and its influence on the degree of localization,
as measured by the information entropy, can be obtained in this system.
In Fig. 7  the quantity $| \mbox{Tr}\;J(\epsilon) - 2 |$, an indicator of 
the stability of the channel orbit, and the information entropy
for the first  2000 states are plotted against $\epsilon$ 
(here the monodromy matrix is for the full Poincar\'{e} map). Neglecting
the envelope of the information entropy, which follows the predictions of
random matrix theory \cite{ssa}, the entropy 
of channel localized states show remarkable oscillations that strongly 
correlate with 
the stability oscillations  of the channel orbit. 

We note from Fig. 7 that the dark circles
corresponding to $ \mbox{Tr}\;J(\epsilon) = 2$ are points of
pitchfork bifurcation at which channel orbit loses stability, and this
correlates strongly with the entropy minima of the channel states. We
may note that when the orbit gains stability the entropy is not 
a minimum although Gutzwiller's trace formula breaks down here as well;
and Berry's scarring amplitude formula \cite{ber} may diverge.
Thus the entropy minima must also have to do with the local structure 
around the periodic orbit and not depend only on the stability matrix 
of the scarring orbit. 

The study of the simplest scarred states, the channel localized ones,
have thus shown interesting exponential localization properties as well
as strong correlation of the degree of localization with the local 
structure of the scarring orbit including its stability. Other states that 
share some of these properties are being actively studied. The scarred state 
noted in the experiment in ref.\cite{wil} also has been shown to obey an 
underlying approximate WKB like eigenvalue  formula \cite{wil}, 
and it may well be possible to experimentally observe some 
of the phenomena noted in this Letter; either by such experiments or possible
microwave cavity experiments \cite{sri}.

\end{multicols}

\begin{figure}[h]
\hspace*{.5in}\psfig{figure=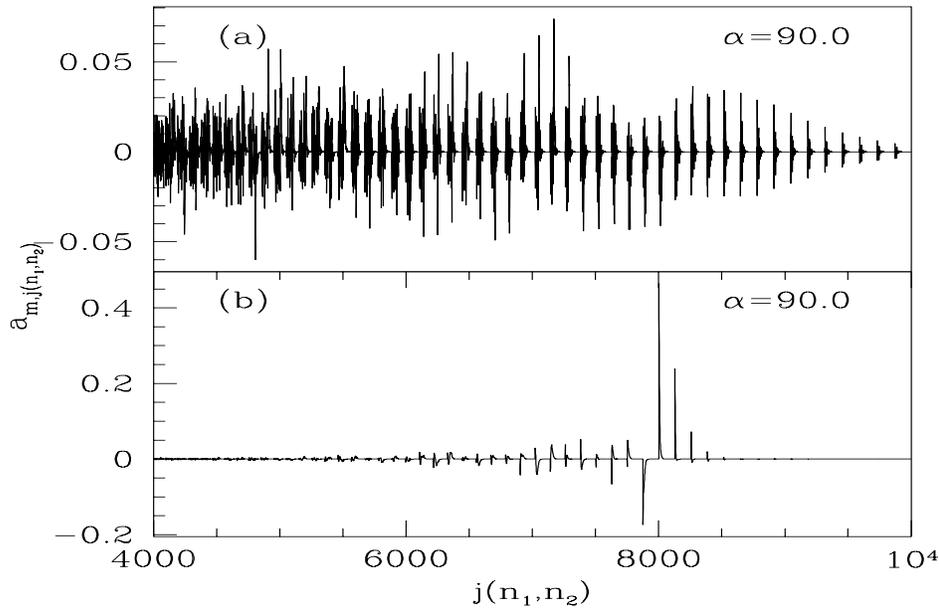,height=3.5in,width=5in}
\vspace{0.4cm}
\caption{
Significant expansion coefficients of two states from the 
same spectrum for the quartic oscillator;  (a) a typical state, 
number 1971 from the ground state and 
(b) state number 1774, a channel localized state.}
\end{figure}

\begin{figure}[h]
\hspace*{.5in}\psfig{figure=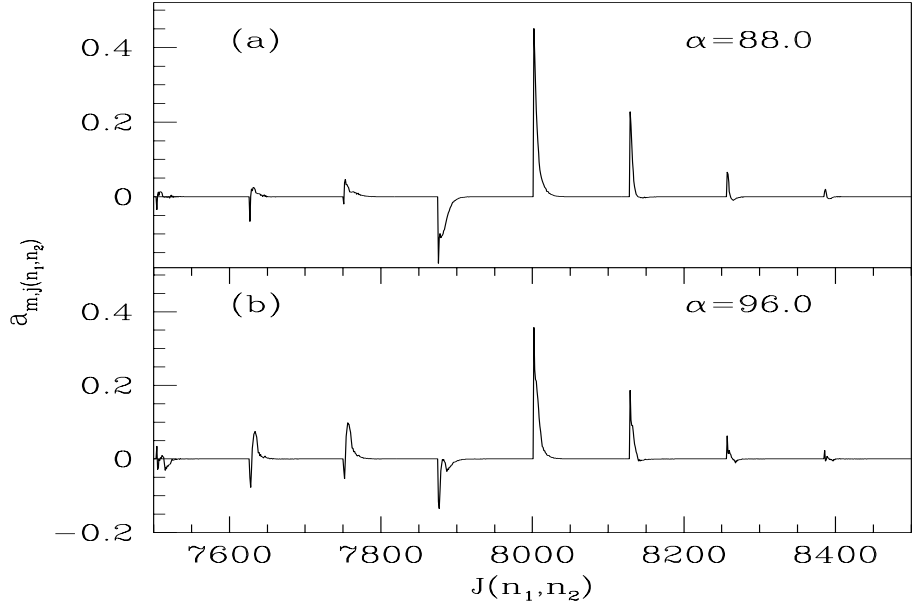,height=3.5in,width=5in}
\vspace{0.4cm}
\caption{(a) Significant expansion coefficients
of two channel localized states at 
different parameter values, (a) state number 1786  and (b) state number 1740.
The principal peak $N=252$ (see text) for these states and those in Fig. 3. }
\end{figure}

\begin{figure}[h]
\hspace*{.5in}\psfig{figure=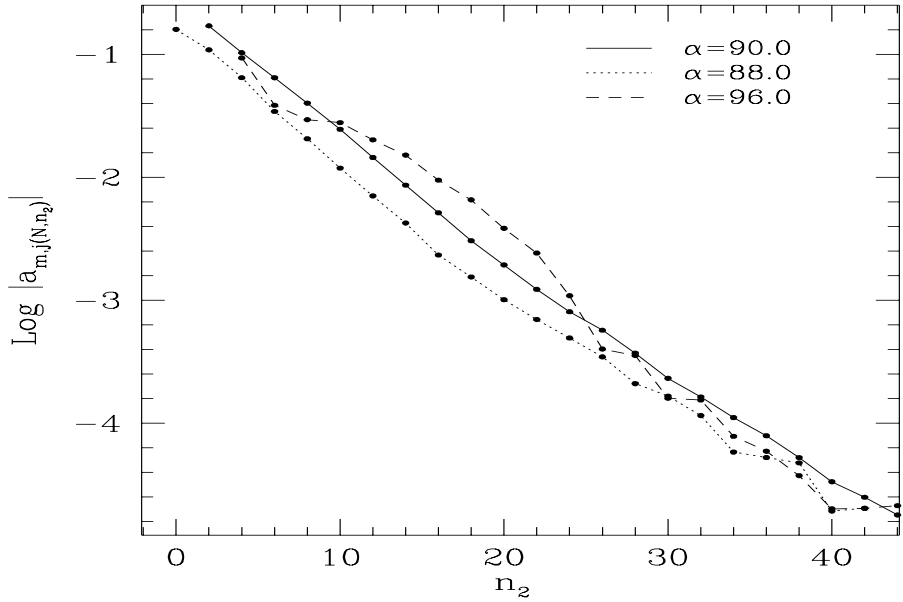,height=3.5in,width=5in}
\vspace{0.4cm}
\caption{Exponential localization in states 1786, 1774 and
1740 at
$\alpha=88.0, 90.0$ and $96.0$ respectively. The values of $n_{2}$ for
each curve is shifted for clarity.}
\end{figure}

\begin{figure}[h]
\hspace*{.5in}\psfig{figure=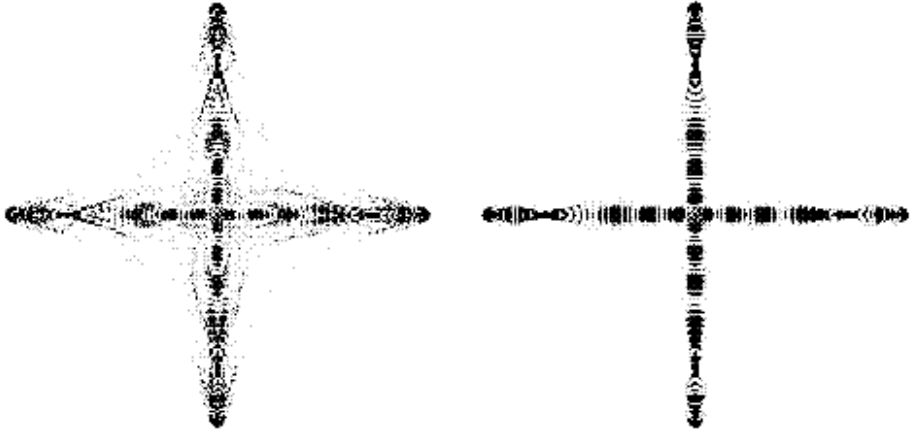,height=5in,width=5in}
\vspace{-3cm}
\caption{Configuration space channel localized
wavefunction densities  for
states 1740 ($\alpha=96.0$, left) and 1774 ($\alpha=90.0$, right)}.
\end{figure}

\begin{figure}[h]
\hspace*{.5in}\psfig{figure=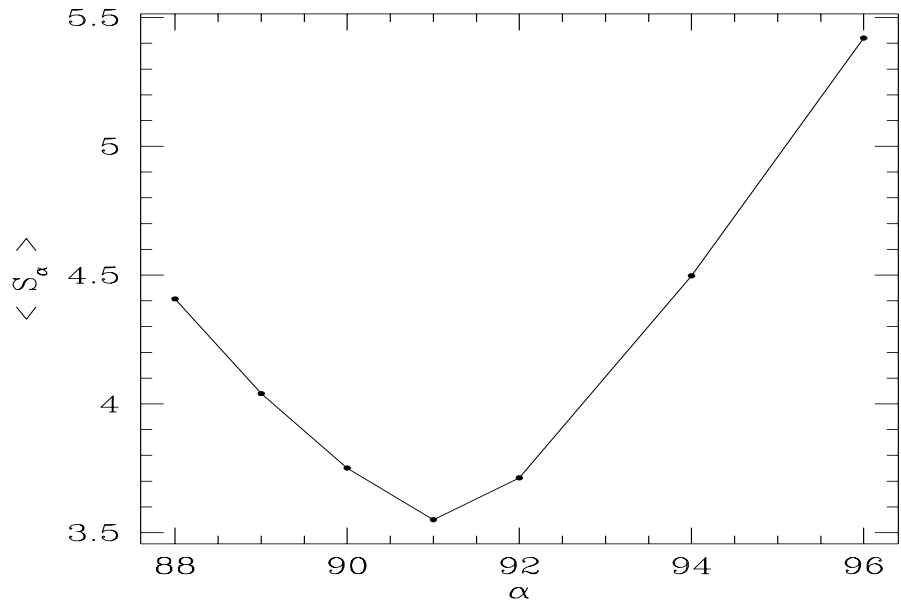,height=3.5in,width=5in}
\vspace{0.4cm}
\caption{Averaged information entropy plotted against
$\alpha$ for the quartic oscillator.}
\end{figure}

\begin{figure}[h]
\hspace*{.5in}\psfig{figure=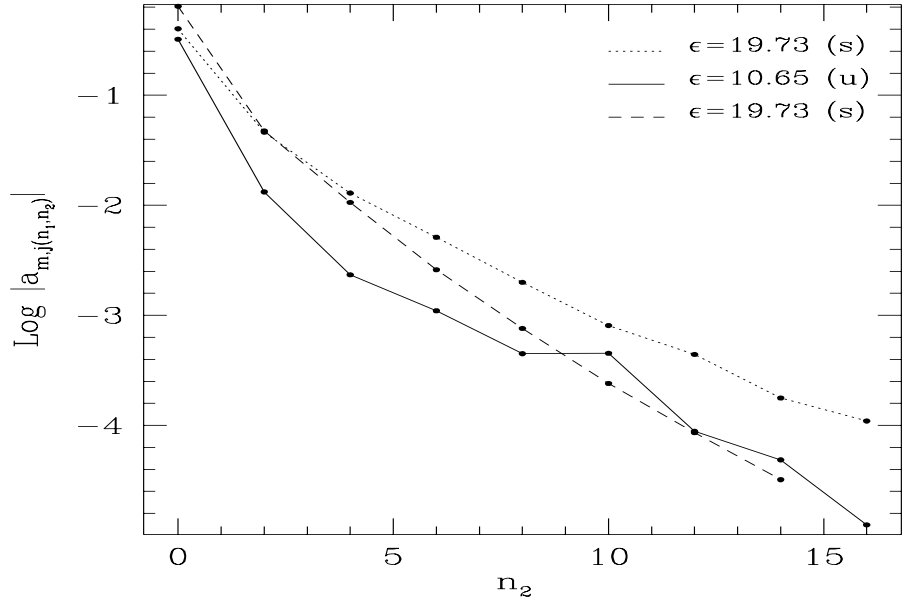,height=3.5in,width=5in}
\vspace{0.4cm}
\caption{Exponential localization for three  channel localized
states of the perturbed oscillator. $u$ and $s$ represent unstable
and stable channel orbits.  $n_{1}$ for
the three curves above are not the same.} 
\end{figure}

\begin{figure}[h]
\hspace*{.5in}\psfig{figure=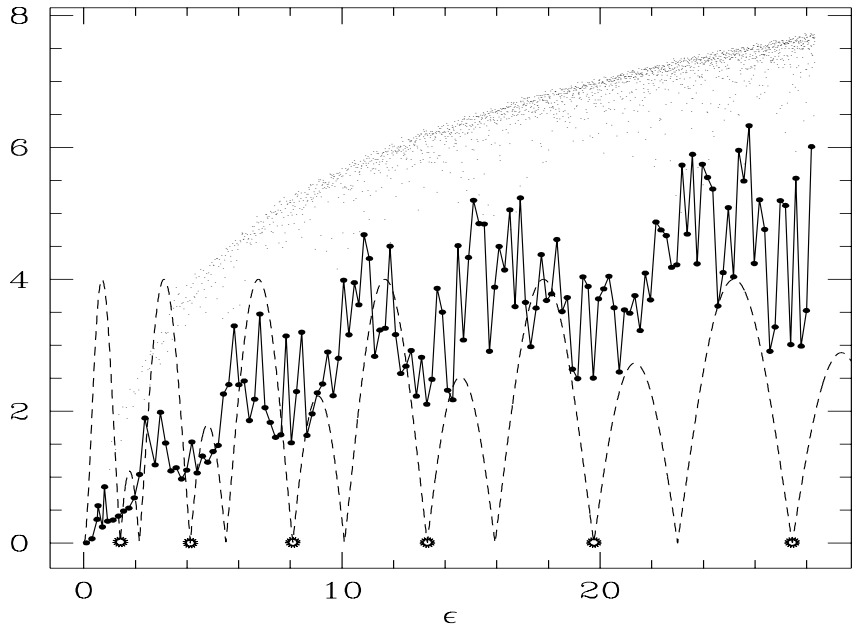,height=3.5in,width=5in}
\vspace{0.4cm}
\caption{Entropy of the first 2000 states of the
perturbed oscillator. The solid line connects the channel localized
states only while the dashed line is $|\mbox{Tr}(J(\epsilon))-2|$.
See text for details.}
\end{figure}

\end{document}